\documentclass[aps,prl,twocolumn,superscriptaddress,nofootinbib]{revtex4-2}
\usepackage{bm}
\usepackage{bbm}
\usepackage{epsfig,graphics,graphicx}
\usepackage{amsmath,amssymb,amsfonts}
\usepackage{color}
\usepackage{epstopdf}
\usepackage{slashed}
\usepackage[
colorlinks=true,
linkcolor=black,
breaklinks=true,
urlcolor=blue,
citecolor=fuchsia]{hyperref}

\usepackage[font=footnotesize, labelfont=bf]{caption,subcaption}
\usepackage{booktabs}
\usepackage{dcolumn}
\usepackage{makecell}
\usepackage{multirow}
\newcolumntype{d}{D{.}{.}{5}}

\definecolor{blueviolet}{rgb}{0.541, 0.169, 0.886}
\definecolor{fuchsia}{rgb}{1.0, 0, 1.0}
\newcommand{\be}{\begin{equation}}
	\newcommand{\ee}{\end{equation}}
\newcommand{\ba}{\begin{eqnarray}}
	\newcommand{\ea}{\end{eqnarray}}

\newcommand{\abs}[1]{\lvert#1\rvert}

\begin{document}
\title{New insights into the nucleon's electromagnetic structure}
\author{Yong-Hui Lin}
\affiliation{Helmholtz Institut f\"ur Strahlen- und Kernphysik and Bethe Center
   for Theoretical Physics, Universit\"at Bonn, D-53115 Bonn, Germany}
\author{Hans-Werner Hammer}
\affiliation{Technische Universit\"{a}t Darmstadt, Department of Physics, 64289 Darmstadt, Germany}
\affiliation{ExtreMe Matter Institute EMMI and Helmholtz Forschungsakademie Hessen f\"ur FAIR (HFHF), GSI Helmholtzzentrum f\"{u}r Schwerionenforschung GmbH,
64291 Darmstadt, Germany}
\author{Ulf-G. Mei{\ss}ner}
\affiliation{Helmholtz Institut f\"ur Strahlen- und Kernphysik and Bethe Center
   for Theoretical Physics, Universit\"at Bonn, D-53115 Bonn, Germany}
\affiliation{Institute for Advanced Simulation and Institut f{\"u}r Kernphysik,
            Forschungszentrum J{\"u}lich, D-52425 J{\"u}lich, Germany}
\affiliation{Tbilisi State University, 0186 Tbilisi, Georgia}
\date{\today}
%
\begin{abstract}
  We present a combined analysis of the electromagnetic form factors of the
  nucleon in the space- and timelike regions using dispersion theory.
  Our framework provides a consistent description of the experimental data
  over the full range of momentum transfer, in line with the
  strictures from analyticity and unitarity. The statistical
  uncertainties of the extracted form factors are estimated using the
  bootstrap method, while systematic errors are
  determined from variations of the spectral functions.
  We also perform a high-precision extraction of the 
  nucleon radii and find good agreement with previous analyses
  of spacelike data alone. For the proton charge radius, we
  find $r_E^p = 0.840^{+0.003}_{-0.002}{}^{+0.002}_{-0.002}$~fm,
  where the first error is statistical and the second one is systematic.
  The Zemach radius and third moment are in agreement with
  Lamb shift measurements and hyperfine splittings.
  The combined data set of space- and timelike data disfavors
  a zero crossing of  $\mu_p G_E^p/G_M^p$ in the spacelike region.
  Finally, we discuss the status and perspectives of modulus and
  phase of the form factors in the timelike region in the context
  of future experiments as well as the onset of perturbative QCD.
\end{abstract}
\maketitle

Our everyday matter consists of electrons, protons, and neutrons,
with the latter two accounting for essentially all of its mass.
While the electron is an elementary particle, protons ($p$)
and neutrons ($n$), which are collectively referred to
as nucleons ($N$), arise from the complicated strong
interaction dynamics of quarks and gluons in
Quantum Chromodynamics (QCD)~\cite{Wilson:1974sk,Wilczek:2012ab}.
The electromagnetic (em) form factors of the nucleon describe the
structure of the nucleon as seen by an electromagnetic probe.
As such, they provide a window on the strong interaction dynamics
in the nucleon over a large range of momentum transfers. For recent reviews
see, e.g. Refs.~\cite{Denig:2012by,Pacetti:2014jai,Punjabi:2015bba}.
Moreover, they are an important ingredient in the description of
a wide range of observables ranging from the Lamb shift in atomic
physics \cite{Pohl:2010zza,Beyer:2017gug,Fleurbaey:2018fih,Bezginov:2019mdi}
over the strangeness content of the
nucleon~\cite{Armstrong:2012bi,Maas:2017snj}
to the em structure and reactions of atomic nuclei
\cite{Bacca:2014tla,Phillips:2016mov,Krebs:2020pii}.
At small momentum transfers, they are sensitive
to the gross properties of the nucleon like the charge and magnetic moment
as well as the radii. At large momentum transfer, they
probe the  quark substructure of the nucleon as described by QCD.

Most discussions of nucleon structure focus on the so-called spacelike
region which is accessible via the Lamb shift or elastic electron scattering
off the nucleon ($e^- N \to e^- N$), where the four-momentum
transfer to the nucleon is spacelike. However,
crossing symmetry connects elastic electron scattering
to the creation of nucleon-antinucleon
pairs in $e^+ e^-$ annihilation and its reverse reaction ($e^+ e^- \leftrightarrow N\bar{N}$).
Both types of processes are described by the Dirac and Pauli form factors
$F_1$ and $F_2$. They depend on the  four-momentum transfer squared $t$
which is defined in the complex plane.
The experimentally accessible spacelike ($t<0$) and timelike
regions ($t>(2m)^2$, with $m = 938.9$~MeV the nucleon mass)
on the real axis are connected by an analytic continuation.
Experimental data are usually given for the Sachs form factors
$G_E$ and $G_M$, which are linear combinations of $F_1$ and $F_2$
and have a physical interpretation in terms of the distribution of charge
and magnetization, respectively (see Methods for details). Note that in
the timelike region, the form factors are complex-valued functions.

The framework of dispersion theory allows to exploit
this link between the space- and timelike data
through a combined analysis of experimental data in both regions, fully consistent
with the fundamental requirements of unitarity and analyticity.
Building upon our previous analyses of spacelike data only
\cite{Lin:2021umk,Lin:2021umz}, we explore this powerful
connection and highlight its consequences for the nucleon radii, the
behavior of the proton form factor ratio $\mu_pG_E^p/G_M^p$, the
onset of perturbative QCD (pQCD) as well as
the modulus and phase of the form factors in the timelike region.
In particular, we discuss the implications of the timelike data for the
``proton radius puzzle'' \cite{Bernauer:2014cwa},
an apparent discrepancy between the proton radius
extracted from the Lamb shift in muonic hydrogen and the value extracted
from electron scattering and the electronic Lamb shift,
see, e.g., Ref.~\cite{Hammer:2019uab,Karr:2020wgh} for the current status of this
puzzle.  It is also important to stress that in the
timelike region, the measured cross section data show an interesting and unexpected oscillatory
behaviour \cite{BaBar:2013ves,BESIII:2021rqk}.

The matrix element for the creation of a nucleon-antinucleon pair
from the vacuum by the em vector current $j_\mu^{\rm em}$
can be expressed as:
\begin{eqnarray}
&& \langle N(p') \overline{N}(\bar{p}') | j_\mu^{\rm em}(0) | 0 \rangle
\nonumber\\
&&\quad = \bar{u}(p') \left[ F_1 (t) \gamma_\mu +i\frac{F_2 (t)}{2 m} \sigma_{\mu\nu}
(p'+\bar{p}')^\nu \right] v(\bar{p}')\,,
\label{eqJ}
\end{eqnarray}
where $p'\,,\bar{p}'$ are the momenta of the nucleon-antinucleon pair
and $t=(p'+\bar{p}')^2 > 0$ is the four-momentum transfer squared.
The analytic structure of this matrix element can be discerned  
by using the optical theorem. Inserting a complete set of
intermediate states $|\lambda\rangle$, one finds \cite{Chew:1958zjr,Federbush:1958zz}
\begin{eqnarray}
&&{\rm Im}\, \langle N(p') \overline{N}(\bar{p}') | j_\mu^{\rm em}(0) | 0 \rangle
 \propto \sum_\lambda
 \langle N(p') \overline{N}(\bar{p}') | \lambda \rangle
 \nonumber\\
&&\quad\times
\langle \lambda | j_\mu^{\rm em} (0) | 0 \rangle \,v(\bar{p}')
\,\delta^4(p'+\bar{p}'-p_\lambda)\,.
\label{spectro}
\end{eqnarray}
Thus the imaginary part of the form factors can be related
to the matrix element for creation of the intermediate states and
the matrix element for scattering of the intermediate states into
a $N\bar{N}$ pair.
The states  $|\lambda\rangle$ must carry the same quantum numbers as
the current $j^{\rm em}_\mu$, i.e., 
$I^G(J^{PC})=0^-(1^{--})$ for
the isoscalar component and $I^G(J^{PC})=1^+(1^{--})$ for the
isovector component. Here, $I, G, J, P,$ and $C$ denote the isospin, G-parity, spin, parity and charge conjugation quantum numbers, in order.
For the isoscalar $(s)$ part  with $I=0$ the lowest mass states are: $3\pi$,
$5\pi$, $\ldots$; for the isovector $(v)$ part  with $I=1$ they
are: $2\pi$, $4\pi$, $\ldots$.
Associated with each intermediate state is a branch
cut starting at the corresponding threshold in $t$ and running to
infinity.

This analytic structure can be exploited to reconstruct the full form factor
from its imaginary part given by Eq.~(\ref{spectro}).
Let $F(t)$ be a generic symbol for one of the nucleon form
factors $F_1$ and $F_2$. Applying Cauchy's theorem to $F(t)$, we obtain
a dispersion relation,
\begin{equation}
  F(t) =\lim_{\epsilon \to 0^+}
  \frac{1}{\pi} \, \int_{t_0}^\infty \frac{{\rm Im}\, 
F(t')}{t'-t-i\epsilon}\, dt'\, ,
\label{emff:disp} 
\end{equation}
which relates the form factor to an integral over its imaginary part
${\rm Im}\, F$. Of course, the derivation assumes that the integral
in Eq.~(\ref{emff:disp}) converges. This is the case for our
parametrization of ${\rm Im}\, F$ (see Methods).

The longest-range, and therefore at low momentum transfer most 
important continuum contribution to the spectral function ${\rm Im}\, F(t)$
comes from the $2\pi$ intermediate state
which contributes to the isovector form factors~\cite{Frazer:1960zzb}.
The $\rho$ appears naturally as a resonance in the $2\pi$ continuum with
a prominent continuum enhancement on its left wing.
A novel and very precise  calculation of this contribution 
has recently been performed in Ref.~\cite{Hoferichter:2016duk} including
the state-of-the-art pion-nucleon scattering amplitudes from dispersion
theory~\cite{Hoferichter:2015hva}.
In the isoscalar channel, the nominally longest-range $3\pi$ contribution
shows no such enhancement and is well accounted for by the $\omega$ pole
\cite{Bernard:1996cc,Kaiser:2019irl}. The most important isoscalar continuum
contributions are the $K\bar{K}$ \cite{Hammer:1998rz,Hammer:1999uf}
and  $\rho\pi$ continua \cite{Meissner:1997qt} in the mass region
of the $\phi$, which is also included as an explicit pole.
The remaining contributions to the spectral function
above $t\approx 1$~GeV can be parameterized by effective
vector meson poles which are fitted to the form factor and cross section data.
Since the analytical continuation from the space- to the timelike region
is, strictly speaking, an ill-posed problem,
the general strategy is to include as few effective poles as possible
to describe the data in order to improve the stability
of the fit \cite{SabbaStefanescu:1978hvt}.

The number of parameters is reduced by applying various constraints.
The asymptotic behavior of the form factors at large spacelike momentum
transfer is constrained by perturbative QCD \cite{Lepage:1980fj}.
The power behavior of the form factors leads to superconvergence relations
which reduce the number of fit parameters.
Moreover, we constrain the fits to reproduce the high-precision determination
of the neutron charge radius squared based on a chiral effective field theory
analysis of electron-deuteron scattering~\cite{Filin:2020tcs},
$\langle r_n^2\rangle = -0.105^{+0.005}_{-0.006}~{\rm fm}^2$.
All other radii are extracted from the analysis of the data.
A detailed discussion of the spectral function is given in Methods.

The data sets included in our fits are listed in Table~\ref{tab:dbase}.
The first five rows contain spacelike data obtained in elastic
electron scattering. Explicit references can be found in
the review \cite{Lin:2021umz}. In the last four rows we list the 
timelike data sets (see Methods for explicit references).
The total number of data points in our analysis is 1753.
\begin{table}[ht!]
	\vspace{2mm}
	\centering  
	\begin{tabular}{|c|c|c|}
		\hline
		Data type                 &  range of $|t|$ [GeV$^2$] & \# of data   \\
		\hline
		$\sigma(E,\theta)$, PRad  &  $0.000215 - 0.058$  & 71        \\
		$\sigma(E,\theta)$, MAMI  &  $0.00384  - 0.977$     & 1422      \\
		$\mu_p G_E^p/G_M^p$, JLab  &  $1.18 - 8.49$     & 16        \\
		$G_E^n$, world            &  $0.14 - 1.47$     & 25        \\
		$G_M^n$, world            &  $0.071- 10.0 $     & 23        \\
		$|G_{\rm eff}^p|$, world            &  $3.52- 20.25 $     & 153        \\
		$|G_{\rm eff}^n|$, world            &  $3.53- 9.49 $     & 27        \\
		$\abs{G_E/G_M}$, BaBar            &  $3.52- 9.0 $     & 6        \\
		$d\sigma/d\Omega$, BESIII            &  $1.88- 1.95$     & 10        \\
		\hline
	\end{tabular}
	\caption{Data sets included in the combined space- and timelike fits.
        See Ref.~\cite{Lin:2021umz} and Methods for explicit references.}
	\label{tab:dbase}
	\vspace{-3mm}
\end{table}

We have started with fits to the timelike data only.
Since the separation of $G_E$ and $G_M$ requires
differential cross sections,
most timelike data are given for the so-called effective form factor
\begin{align}
\left|G_{\rm eff}\right| \equiv \sqrt{\frac{|G_E|^2+\xi|G_M|^2}{1+\xi}}~,
\label{geff}
\end{align}
with $\xi=t/(2m^2)$.
However, there are also some data for the ratio $\abs{G_E/G_M}$ and
some differential cross section  data from BaBar and BESIII. 
The phase of the ratio $G_E/G_M$ has not been measured. It turns out that
a certain number of broad poles above threshold is needed to get 
a good description of the  timelike data. These poles generate the imaginary
part of the form factors above the two-nucleon threshold and are required
to describe the observed oscillatory behavior of the form factors
from BaBar and  BESIII.
With $3s+3v$ below-threshold narrow poles and $3s+3v$ above-threshold broad
poles, we were able to obtain a good fit to the data
with $\chi^2/{\rm d.o.f}=0.638$. In particular, the visible strong enhancement
of the proton and the neutron timelike form factor (after subtraction of
the electromagnetic final-state interaction in the proton case), first seen
by the PS170 collaboration at LEAR~\cite{Bardin:1994am},  is also described
in this framework.

In the next step, we include the spacelike data and aim for
a consistent analysis of both types of data.
We explicitly enforce a decreasing behavior of $G_M^n/(\mu_n G_{\rm dip})$
$\mu_p G_E^p/G_M^p$ at large $|t|$ in the spacelike region
in order to get a good description over the full range of momentum
transfers. Moreover, the weight of the timelike ratio $\abs{G_E/G_M}$
data from BaBar is increased by a factor of 10 so as to make its
contribution to the total $\chi^2$ that is highly suppressed by the
large uncertainties more sizable.  

\begin{figure}[htbp]
	\centering
		\includegraphics[width=0.47\textwidth]{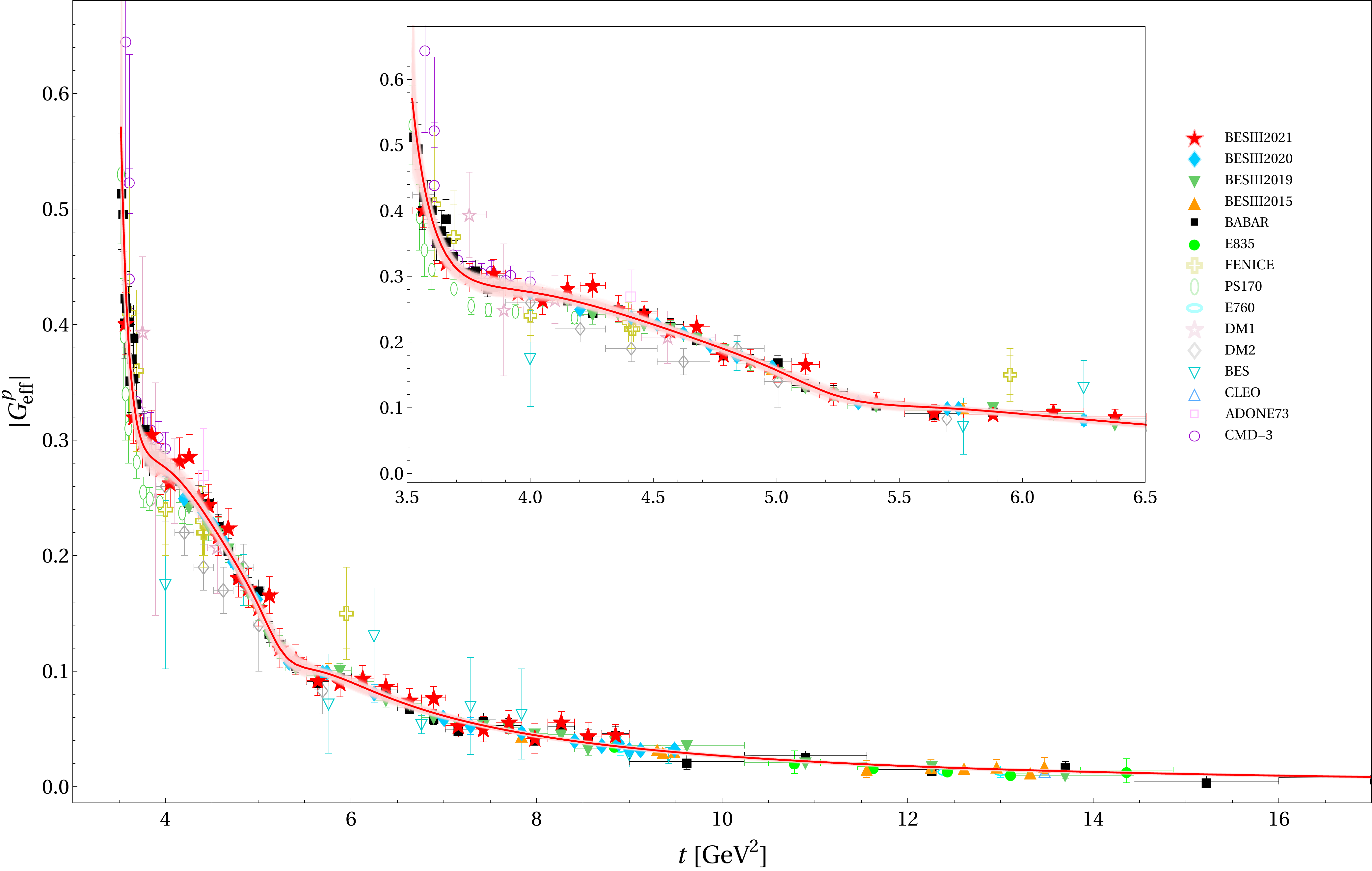}
		\includegraphics[width=0.47\textwidth]{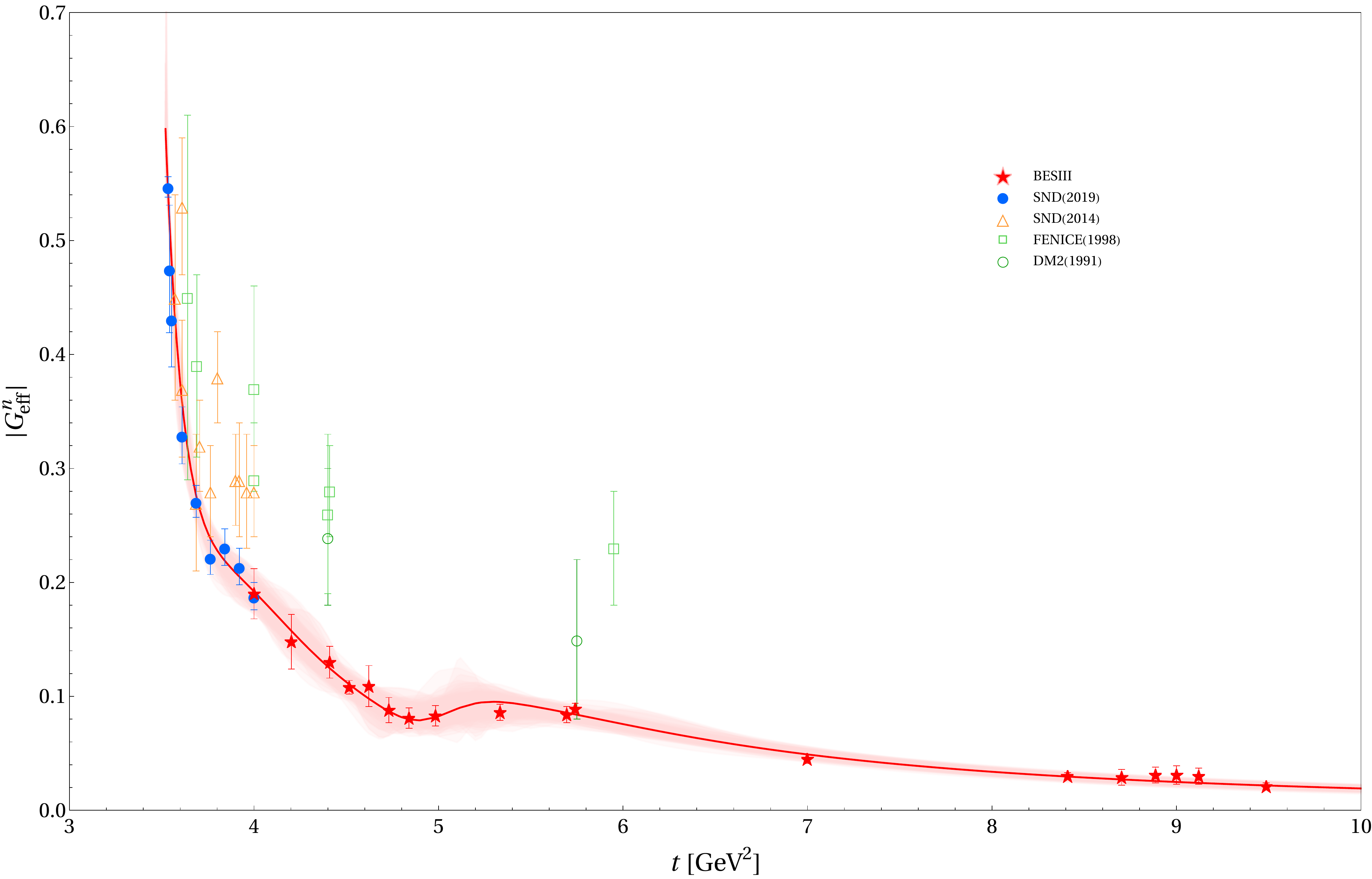}
       		\caption{Complete fit to space- and timelike data
                  with bootstrap error (shaded band)
                  compared to data for $|G_{\rm eff}|$
                  of the proton (upper panel) and neutron (lower panel).
                  Fitted data are depicted by closed symbols; data given
                  by open symbols are shown for comparison
                  only (see Methods for explicit references).}
			\label{Fig: geffp}
	\end{figure}

In Fig.~\ref{Fig: geffp}, we show our best fit compared to
the experimental data for $|G_{\rm eff}|$ of the proton (upper panel)
and the neutron (lower panel). We obtain a good description of
the timelike data for $|G_{\rm eff}|$.
The prominent oscillations in $|G_{\rm eff}|$
between the threshold at $t = 4m^2$ and $t\approx 6$~GeV$^2$
are reproduced by the effective broad poles above threshold.
These poles also generate the imaginary part
of the form factors in the physical region. 
Alternatively, these structures can also be generated by including
contributions from triangle diagrams with $\Delta\bar{\Delta}$
and $(\Delta\bar{N}+{\rm h.c.})$ intermediate states, see, e.g.,
Ref.~\cite{Lorenz:2015pba}. In principle, these contributions
are fixed. However, the corresponding coupling constants are poorly
known and a perturbative treatment of these contributions
is questionable. For further discussion, see Ref.~\cite{Bianconi:2015owa}.

The quality of the fit to the spacelike data is comparable to our
previous fits of spacelike data only \cite{Lin:2021umk,Lin:2021umz}.
We obtain $\chi^2/{\rm d.o.f}=1.223$ for the full data set,
$\chi^2/{\rm d.o.f}=1.063$ for the timelike data, and 
$\chi^2/{\rm d.o.f}=1.297$ for the spacelike data.
Thus it is warranted to extract the nucleon radii from our
combined fit, which has a larger data base than spacelike
only fits. We obtain the radii
\begin{eqnarray}
r_E^p &=& 0.840^{+0.003}_{-0.002}{}^{+0.002}_{-0.002}\,{\rm fm}, \nonumber\\
r_M^p &=& 0.849^{+0.003}_{-0.003}{}^{+0.001}_{-0.004}~{\rm fm}, \nonumber\\
r_M^n &=& 0.864^{+0.004}_{-0.004}{}^{+0.006}_{-0.001}~{\rm fm},
%
\label{eq:radii_fin}
\end{eqnarray}
where the first error is statistical (based on the bootstrap procedure
explained in Methods) and the second one is systematic
(based on the variations in the spectral functions, see Methods).
These values are in good agreement with previous high-precision analyses
of spacelike data alone \cite{Lin:2021umk,Lin:2021umz} and
have comparable errors.
For the Zemach radius $r_z$ and the third Zemach
  moment $\langle r^3 \rangle_{(2)}$ (see Methods), we obtain
  \begin{eqnarray}
r_z &=& 1.054^{+0.003}_{-0.002}{}^{+0.000}_{-0.001}\,{\rm fm}, \nonumber\\
\langle r^3\rangle_{(2)} &=& 2.310^{+0.022}_{-0.018}{}^{+0.014}_{-0.015}~{\rm fm}^3.
\label{eq:radii_zemach}
  \end{eqnarray}
These values are in good agreement with Lamb shift and hyperfine splittings in
muonic hydrogen \cite{Antognini:2013txn}.

\begin{figure}[htbp]
	\centering
		\includegraphics[width=0.47\textwidth]{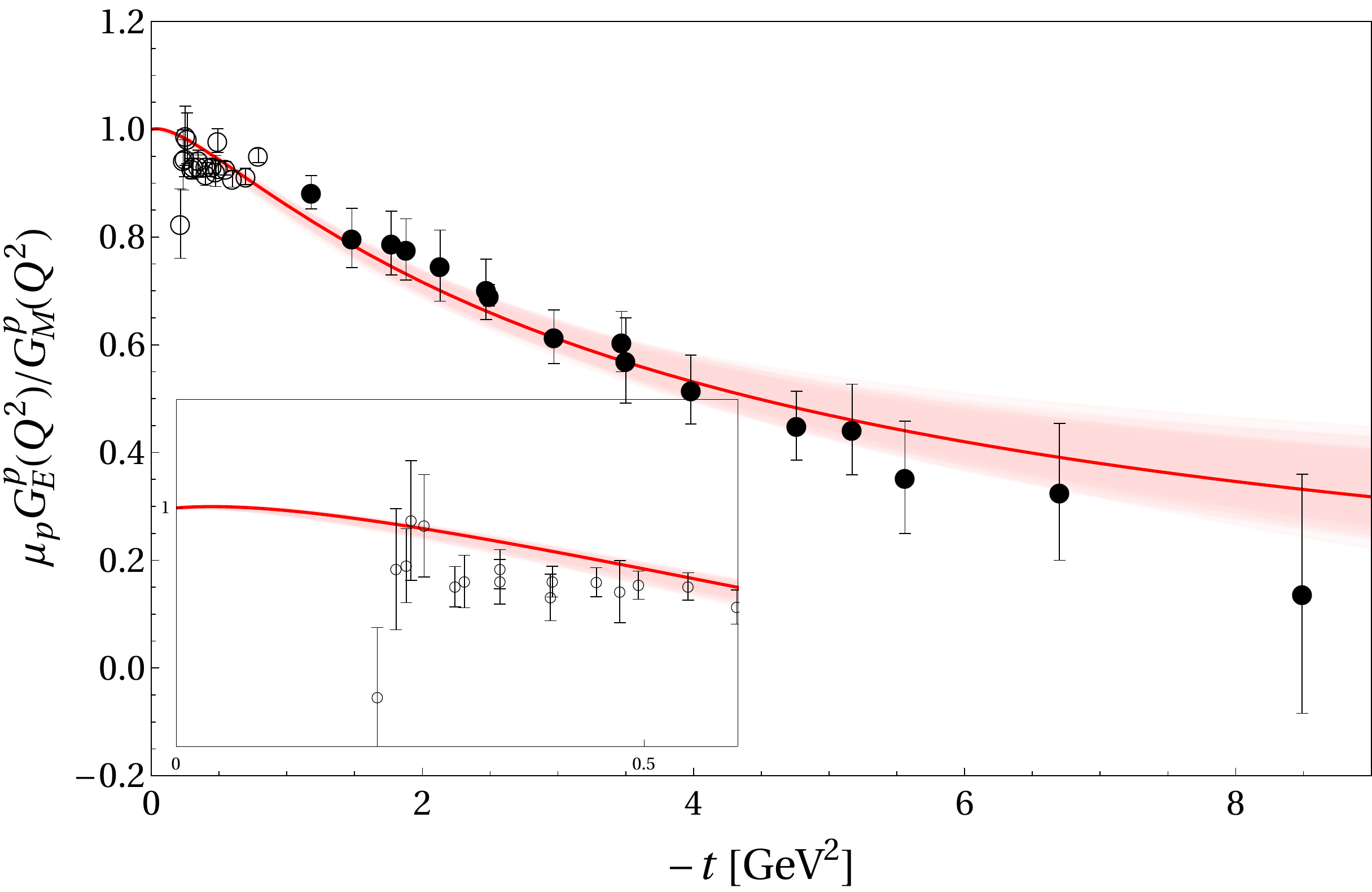}
		\caption{Complete fit to space- and timelike data
                  with bootstrap error (shaded band) compared to JLab
                  data for $\mu_p G_E^p/G_M^p$
                  at spacelike momentum transfer.
                  Fitted data are depicted by closed symbols. The data
                  for $|t|<1$~GeV$^2$
                  (open symbols, see also the inset) are
                  shown for comparison only. For references to the data,
                  see Methods.
			\label{Fig: geffpR}
		}
\end{figure}

Another interesting question in the spacelike region concerns
the behavior of the form factor ratio $\mu_p G_E^p/G_M^p$
for intermediate momentum transfer. Some measurements suggest
a zero crossing of this ratio around $t \approx -10$~GeV$^2$
\cite{Arrington:2011kb}. In Fig.~\ref{Fig: geffpR}, we compare
our fit to the experimental data for $\mu_p G_E^p/G_M^p$.
While we obtain a good description of the data, a zero crossing
is disfavored by the combined analysis of space- and timelike
data. Thus, data at higher momentum transfer than shown in the figure
are required to settle this issue. We further remark that as in the
earlier fits to the spacelike data only, the onset of perturbative
QCD barely sets in at the highest momentum transfers probed.

\begin{figure}[htbp]
	\centering
		\includegraphics[width=0.47\textwidth]{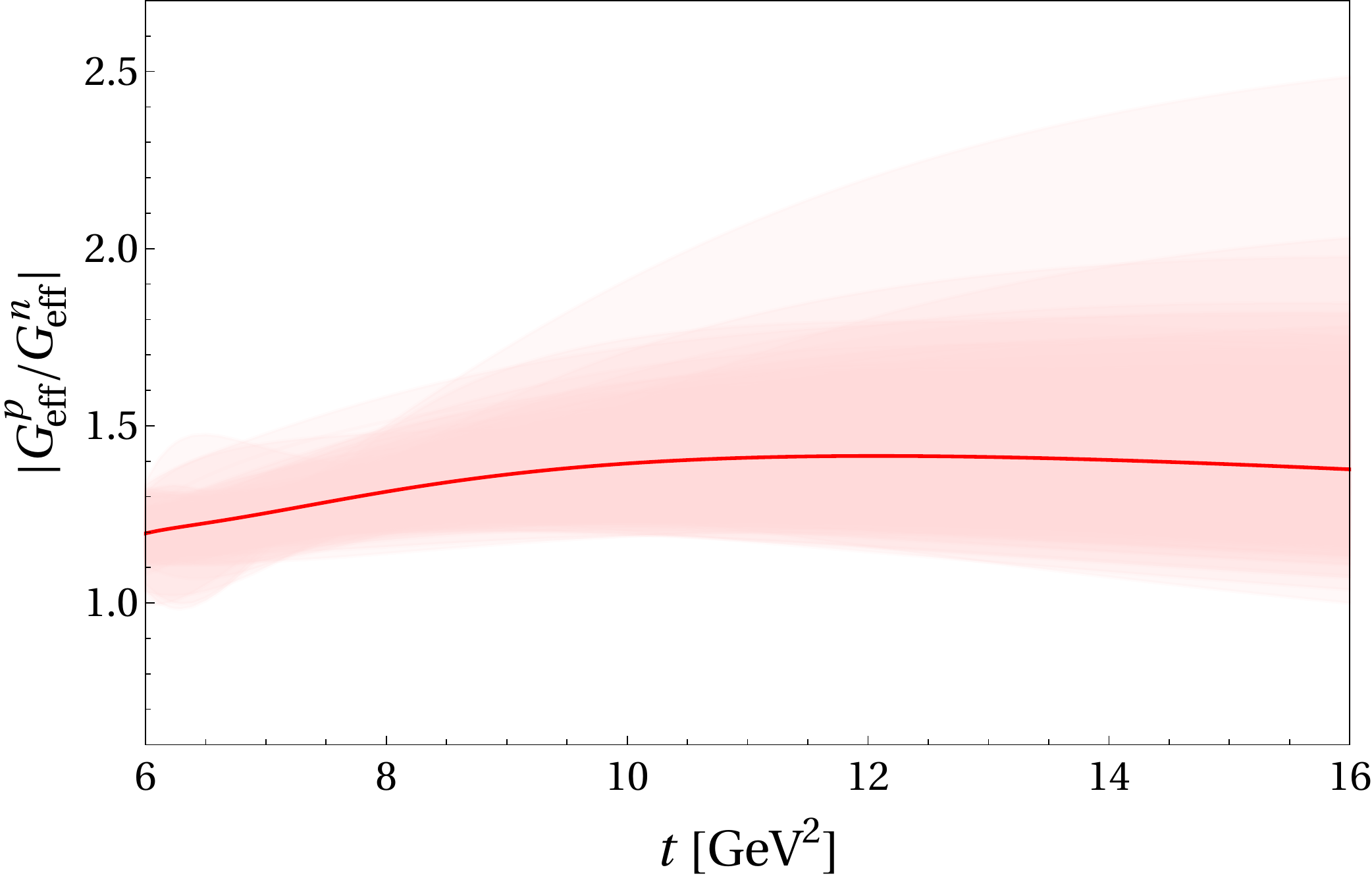}
		\caption{Form factor ratio $|G^p_{\rm eff}(t)/G^n_{\rm eff}(t)|$ in the timelike
                  region for the best fit with the bootstrap uncertainties
                  indicated by the shaded band.
			\label{Fig:ratgeffp}
		}
\end{figure}

Based on quark counting rules \cite{Lepage:1980fj},
the form factor ratio $|G^p_{\rm eff}(t)/G^n_{\rm eff}(t)|$
should approach a constant as $t\to \infty$ in the timelike region.
We show our result for this ratio in  Fig.~\ref{Fig:ratgeffp}.
The form factor ratio is constant above $t \simeq 6\,$GeV$^2$ and slightly
larger than one, with sizeable uncertainties for $t>10$~GeV$^2$. 
However, drawing a clear conclusion about the onset of pQCD certainly requires
the separated form factors $G_E$ and $G_M$,
and not just the effective form factor.

\begin{figure}[htbp]
	\begin{center}
		\includegraphics[width=0.47\textwidth]{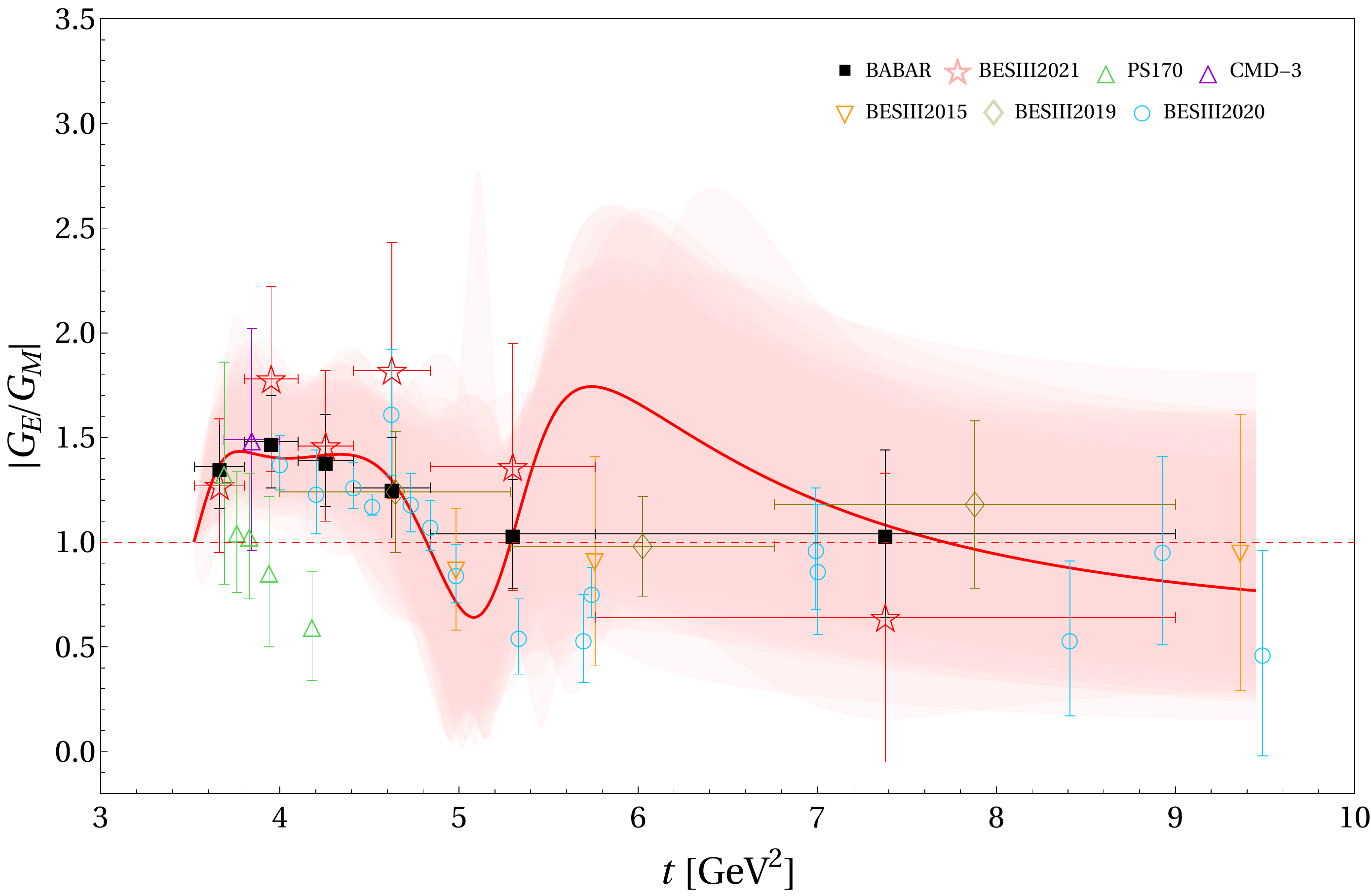}
		\includegraphics[width=0.47\textwidth]{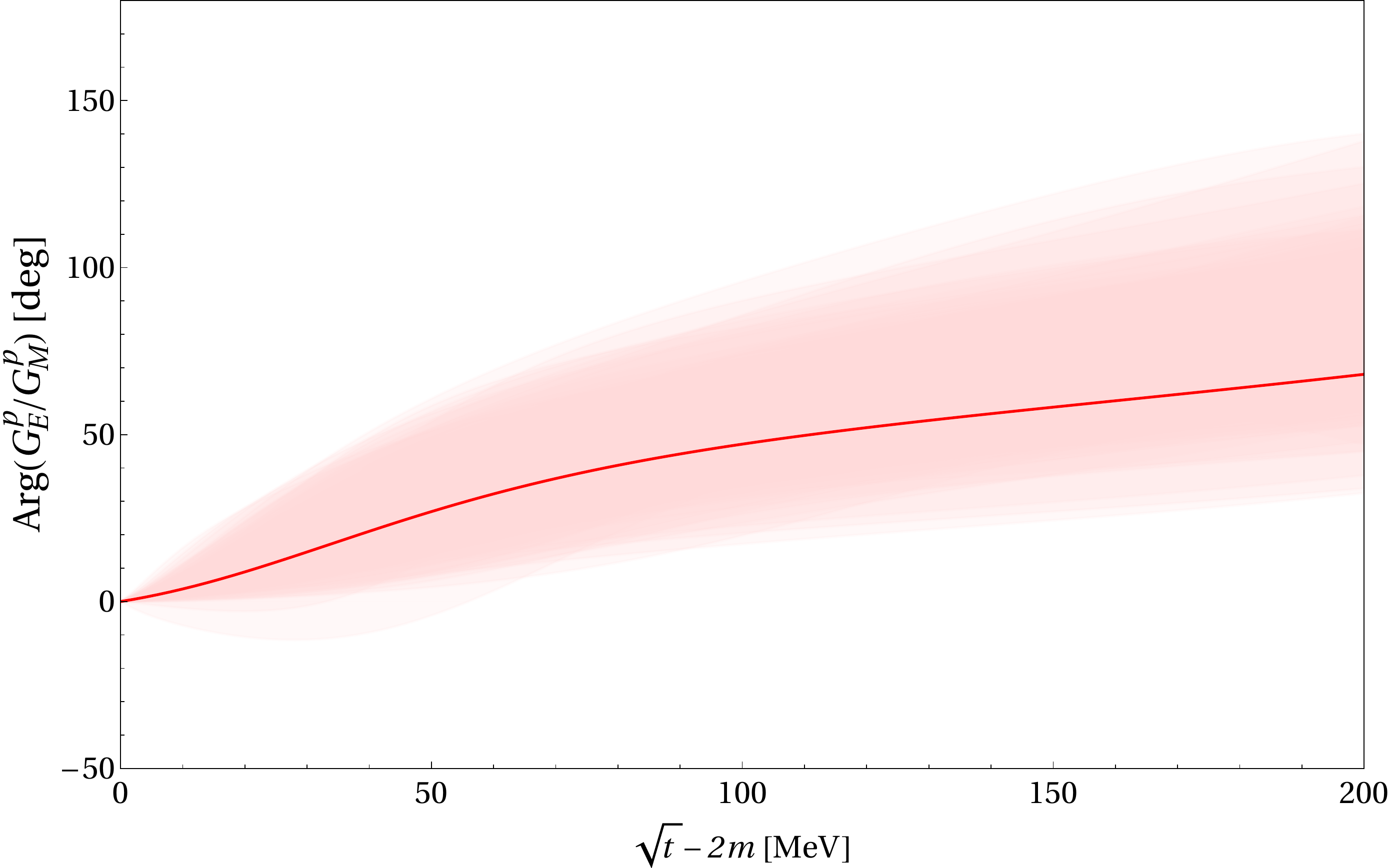}
		\caption{Complete fit to space- and timelike data
                  with bootstrap error (shaded band)
                  compared to proton data for
                  $\abs{G_E/G_M}$ (closed symbols: fitted, open symbols: not fitted)
                  (upper panel) and our
                  prediction for $\arg(G_E/G_M)$
                    (lower panel).
			\label{Fig:rageffp}
		}
	\end{center}
\end{figure}

In addition to $|G_{\rm eff}|$, there are also data on the ratio
$\abs{G_E/G_M}$ and on differential cross sections
for the proton in the timelike region.
The differential cross sections from BESIII  in the lowest energy bin
($t \in [1.877,1.950]$~GeV$^2$) are included in our fit and well described.
The corresponding differential cross section from BaBar are also well described,
when normalized to the total cross section.
In Fig.~\ref{Fig:rageffp}, we compare the fit to the proton data for
$\abs{G_E/G_M}$ and give our prediction for the phase of
$G_E/G_M$. We fit only to the BaBar data for $\abs{G_E/G_M}$
since the BESIII data have much larger error bars.
The modulus $\abs{G_E/G_M}$ is well described by our fit
but the bootstrap errors grow to more than 100\%
at $t\approx 6$~GeV$^2$. The phase $\arg(G_E/G_M)$
is experimentally unrestricted due to the lack of data
and thus has large errors. For energies $\sqrt{t}-2m$
larger than 200~MeV it is essentially unconstrained by our fit.
Future measurements of the phase such as planned
with PANDA at FAIR would be highly valuable to improve this situation
\cite{Dbeyssi:2019ndg}.

In summary, for the first time a consistent picture of the nucleons
electromagnetic structure based on all spacelike and timelike
data from electron scattering and electron-positron annihilation
(and its reversed process) emerges. In particular, the extracted proton charge
radius $r_E^p=0.840\,$fm is small and consistent with earlier
dispersive analyses~\cite{Lin:2021umz} and most recent determinations
from electron-proton scattering as well as the Lamb shift in electronic
and muonic hydrogen (as listed e.g. in Ref.~\cite{Hammer:2019uab}).
The Zemach radius and third moment are in agreement with
Lamb shift measurements and hyperfine splittings in muonic
hydrogen~\cite{Antognini:2013txn}.
Still, there are open
questions related to the onset of pQCD, the behaviour of the
form factor ratio $\mu_p G_E^p/G_M^p$ at intermediate $|t|$ in the spacelike region,
as well as the precise behaviour of this complex-valued ratio in the
timelike region. These issues can only be settled by accurate measurements
combined with precise analyses as in the framework utilized here.

\acknowledgments
{\em Acknowledgements:}~The work of UGM and YHL is supported in
part by the Deutsche Forschungsgemeinschaft (DFG, German Research
Foundation) and the NSFC through the funds provided to the Sino-German Collaborative  
Research Center TRR~110 ``Symmetries and the Emergence of Structure in QCD''
(DFG Project-ID 196253076 - TRR 110, NSFC Grant No. 12070131001),
by the Chinese Aca\-de\-my of Sciences (CAS) through a President's
International Fellowship Initiative (PIFI) (Grant No. 2018DM0034), by the VolkswagenStiftung
(Grant No. 93562), and by the EU Horizon 2020 research and innovation programme,
STRONG-2020 project under grant agreement No. 824093. HWH was supported by the
Deut\-sche Forschungsgemeinschaft (DFG, German
Research Foundation) -- Projektnummer 279384907 -- CRC 1245
and by the German Federal Ministry of Education and Research (BMBF) (Grant
No. 05P18RDFN1).

\appendix

\section{Methods}

\subsection{Definitions}

The matrix element of the  electromagnetic (em) current operator
$j_\mu^{\rm em}$ in the nucleon can be parametrized as
\begin{equation}
\langle N(p') | j_\mu^{\rm em} | N(p) \rangle = \bar{u}(p')
\left[ F_1 (t) \gamma_\mu +i\frac{F_2 (t)}{2 m} \sigma_{\mu\nu}
  q^\nu \right] u(p)\,,
\label{emff:def}
\end{equation}
where $t=(p'-p)^2<0$ is the four-momentum transfer. 
The Dirac and Pauli form factors $F_1$ and $F_2$
are normalized at $t=0$ to the total charge and anomalous magnetic
moment, respectively:
$F_1^p(0) = 1, \; F_1^n(0) = 0, \; F_2^p(0) =  \kappa_p=1.793,$ and $F_2^n(0) =
\kappa_n=-1.913$.
For the dispersion analysis, it is convenient to 
decompose the form factors into isoscalar and isovector parts,
\begin{equation}
F_i^{s/v} = \frac{1}{2} (F_i^p \pm F_i^n) \, , \quad i = 1,2 \,,
\end{equation}
since the intermediate states in Eq.~(\ref{spectro})
have good isospin. The so-called Sachs form factors
\begin{eqnarray}
\label{sachs}
&&G_{E}(t) = F_1(t) +\frac{t}{4m^2} F_2(t) \, , \quad
G_{M}(t) = F_1(t) + F_2(t) \, , \nonumber
\end{eqnarray}
have a more transparent physical interpretation as
Fourier transforms of the charge and magnetization distributions
in the Breit frame, respectively.              

The low-$t$ behavior of the form factors contains information
about the nucleon's size as seen by an electromagnetic probe.
The root mean square radii (loosely called nucleon radii) with
$r \equiv \sqrt{\langle r^2 \rangle}$ are defined via
\begin{equation}
\label{def:r2}
F(t)=F(0)\left[1+t \frac{\langle r^2 \rangle}{6} +\ldots \right]\,,
\end{equation}
where $F(t)$ is a generic form factor. In the case of the electric
and Dirac form factors of the neutron, $G_E^n$ and $F_1^n$, the
expansion starts with the term linear in $t$ and the 
normalization factor $F(0)$ is dropped.

The Zemach radius $r_z$ and third Zemach moment $\left\langle r^3 \right\rangle_{(2)}$ are defined as
\begin{align}
	r_z&=\frac{2}{\pi}\int_{-\infty}^0
	\frac{d t}{t\sqrt{-t}}\left(\frac{G_E(t)G_M(t)}{1+\kappa_p}-1\right)~,\\	
	\left\langle r^3 \right\rangle_{(2)}&=\frac{24}{\pi}\int_{-\infty}^0
	\frac{d t}{t^2\sqrt{-t}}\left(G^2_E(t)-1-\frac{t}3
        \left\langle r^2 \right\rangle_{p}\right)~.
\end{align}

\subsection{Spectral functions}

The spectral function applied in our fits has the following structure:
\begin{eqnarray}
{\rm Im }\,F_i^{s} (t) &=& {\rm Im }\,F_i^{(s,K\bar{K})} (t)
+ {\rm Im }\,F_i^{(s,\rho\pi)} (t) \nonumber \\
&+& \sum_{V=\omega,\phi,s_1,...} \pi a_i^{V}
\delta (M^2_{V}-t)\nonumber \\
&+& \sum_{V=r_{s1},...}  \pi a_i^{V}
\delta (M^2_{V}-i M_V\Gamma_V-t)\, ,
\label{emff:s}
\end{eqnarray}
\begin{eqnarray}
{\rm Im }\,F_i^{v} (t) &=& {\rm Im }\,F_i^{(v,2\pi)} (t)
+ \sum_{V= v_1,...} \pi a_i^{V} \delta (M^2_{V}-t)
\nonumber \\
&+&\sum_{V=r_{v1},...}  \pi a_i^{V}
\delta (M^2_{V}-i M_V\Gamma_V-t)
\,, 
\label{emff:v}
\end{eqnarray}
where $i = 1,2\,$.
It consists of the physical $\omega$ and $\phi$ poles, which have
fixed masses, and both narrow and broad effective vector meson poles.
The masses of all effective poles and the widths of the
broad poles are fitted to the data. Moreover, all vector meson
coupling constants are fitted. 
The $2\pi$, $K\bar{K}$ and $\rho\pi$
continua are determined from other processes and enter as fixed
contributions, see Ref.~\cite{Lin:2021umz} for details.
Our best fit consists of  $5$ narrow poles in the isoscalar channel $(s)$
and 5 narrow poles in the isovector channel $(v)$
below the nucleon-nucleon threshold and $3s+3v$ broad poles above
the threshold.
In addition, there are 33 normalization constants for the MAMI and PRad data
in the spacelike region. These are discussed in detail in
Ref.~\cite{Lin:2021umz}. In total this adds up to 85 parameters.
Including the 11 constraints, namely the 4 for the normalizations
of $F_i^{p/n}(0)$, 6 for the superconvergence relations and 1 for the fixed neutron
charge radius squared, this results in 74 free fit parameters.
The vector meson
parameters of our best fit are listed in Table~\ref{tab:fit-para}.
\begin{table}[ht!]
	\centering
	\begin{tabular}{|l|c|c|c|c|}
		\Xhline{0.5pt}
		$V_{s}$ & $M_V$ &$\Gamma_V$ & $a_1^V$ &$a_2^V$ \\
		\hline
		$\omega$ & $0.783$	&$0$ & $0.701$	& $0.338$	\\
		$\phi$   & $1.019$	&$0$& $-0.526$	& $-0.997$	\\
		$s_1$   & $1.031$	&$0$& $0.422$	& $-2.827$	\\
		$s_2$   & $1.120$	&$0$& $0.122$	& $3.655$	\\
		$s_3$   & $1.827$	&$0$& $0.955$	& $-1.122$	\\
		$r_{s1}$   &$1.903$ &$0.973$ &$-2.653$	& $-1.753$	\\
		$r_{s2}$   &$1.914$ &$0.541$ &$-3.069$	& $2.017$	\\
		$r_{s3}$   &$1.879$ &$0.895$ &$4.953$	& $0.501$	\\
                		\hline \hline
		$V_{v}$ & $M_V$ &$\Gamma_V$ & $a_1^V$ & $a_2^V$  \\
		\hline
		$v_1$ & $1.050$	&$0$ & $0.782$ & $-0.132$	\\
		$v_2$ & $1.323$	&$0$ & $-4.873$ & $-0.645$	\\
		$v_3$ & $1.368$	&$0$ & $3.518$ & $-0.987$	\\
		$v_4$ & $1.462$	&$0$ & $2.243$ & $-3.813$	\\
		$v_5$ & $1.532$	&$0$ & $-1.422$ & $3.668$	\\
		$r_{v1}$   &$2.256$ &$0.239$ &$2.552$	& $-1.217$	\\
		$r_{v2}$   &$2.253$ &$0.245$ &$-1.947$	& $0.551$	\\
		$r_{v3}$   &$2.220$ &$0.362$ &$-0.985$	& $1.061$	\\
		\hline
	\end{tabular}
        \caption{Parameters for best fit to space- and timelike data.
          Masses ($M_V$) and width ($\Gamma_V$) are in GeV while the residua
          $a_{1,2}^V$ are given in GeV$^2$. The broad poles are denoted by the
          symbol $r$.}
        \label{tab:fit-para}
        \end{table}

\subsection{Data basis}

The data set in the spacelike region is the same as in Ref.~\cite{Lin:2021umz}.
In the timelike region, we include the data sets listed in
Table \ref{tab:data}.
\begin{table}[ht!]
	\vspace{2mm}
	\centering  
	\begin{tabular}{|c|c|}
		\hline
		Data type                 &  \rm{Reference}  \\
		\hline
		\multirow{5}{*}{$G_{\rm eff}^p$}
                & BESIII2021~\cite{BESIII:2021rqk}, BESIII2020~\cite{BESIII:2019hdp}, BESIII2019~\cite{BESIII:2019tgo}\\
                & BESIII2015~\cite{BESIII:2015axk},
		BABAR~\cite{BaBar:2013ves}, E835~\cite{E835:1999mlt,Andreotti:2003bt}\\
		& FENICE~\cite{Antonelli:1993vz,Antonelli:1994kq,Antonelli:1998fv}, PS170~\cite{Bardin:1994am}, E760~\cite{E760:1992rvj}\\
		& DM1~\cite{Delcourt:1979ed}, DM2~\cite{Bisello:1983at,DM2:1990tut}, BES~\cite{BES:2005lpy}\\
		& CLEO~\cite{CLEO:2005tiu}, ADONE73~\cite{Castellano:1973wh}, CMD-3~\cite{CMD-3:2015fvi}
		    \\
		\hline
		\multirow{2}{*}{$G_{\rm eff}^n$}
                & BESIII~\cite{BESIII:2021dfy}, SND2019~\cite{Druzhinin:2019gpo}, SND2014~\cite{Achasov:2014ncd}\\
		& FENICE1998~\cite{Antonelli:1998fv}, DM2(1991)~\cite{Biagini:1990nb}
		      \\
		\hline
		\multirow{3}{*}{$\abs{G_{E}/G_{M}}$}
                & BABAR~\cite{BaBar:2013ves}, BESIII2021~\cite{BESIII:2021rqk}, PS170~\cite{Bardin:1994am}\\
		& CMD-3~\cite{CMD-3:2015fvi}, BESIII2015~\cite{BESIII:2015axk},	BESIII2019~\cite{BESIII:2019tgo}\\
                & BESIII2020~\cite{BESIII:2019hdp}
		\\
		\hline
	\end{tabular}
	\caption{Data sets and references for timelike form factors.}
        \label{tab:data}
\end{table}
%

\subsection{Fitting procedure}

The quality of the fits is measured by means of two different $\chi^2$ functions,
$\chi^2_1$ and $\chi^2_2$, which are defined as
\begin{align}
	\chi^2_1 &= \sum_i\sum_k\frac{(n_k C_i - C(t_i,\theta_i,\vec{p}\,))^2}{(\sigma_i+\nu_i)^2}~,
	\label{eq:chi1}\\
	\chi^2_2 &= \sum_{i,j}\sum_k(n_k C_i - C(t_i,\theta_i,\vec{p}\,))[V^{-1}]_{ij}\notag\\
	& \qquad\qquad \times (n_k C_j - C(t_j,\theta_j,\vec{p}\,))~,
	\label{eq:chi2}
\end{align}
where $C_i$ are the experimental data at the points $t_i,\theta_i$ and
$C(t_i,\theta_i,\vec{p}\,)$ are the theoretical value for a given FF parametrization
for the parameter values contained in $\vec{p}$.
For total cross sections and form factor data the dependence on $\theta_i$
is dropped. Moreover, the $n_k$ are normalization
coefficients for the various data sets (labeled by the integer $k$ and only used in the fits to
the differential cross section data in the spacelike region), while $\sigma_i$ and $\nu_i$ are 
their statistical and systematical errors, respectively. The covariance matrix
$V_{ij} = \sigma_i\sigma_j\delta_{ij} + \nu_i\nu_j$.
$\chi^2_2$ is used for those experimental data where statistical and systematical errors are given separately, otherwise 
$\chi^2_1$ is adopted. Furthermore, the $\chi^2$ of each data set is normalized by the number of data points in order to weight the various data sets without bias.
	
As done in Ref.~\cite{Lin:2021umk,Lin:2021umz} the various constraints on the form factors are imposed in a soft way, 
that is, all constraints are implemented as additive terms to the total $\chi^2$ in the following form
\begin{equation}
	\chi^2_{\rm add.} = p\, [x-\langle x\rangle]^2 \,\exp\left(p\, [x-\langle x\rangle]^2\right)~,
\end{equation}
where $\langle x\rangle$ is the desired value and $p$ is a strength parameter,
which regulates the steepness of the exponential well and helps to stabilize the
fits. The fits are performed with {\em MINUIT} \cite{James:1975dr} in Fortran. 

\subsection{Error estimates}

The errors from the fits will be quantified using the bootstrap method.
We simulate a large number of data sets by randomly varying the points in the
original set within the given errors assuming their normal distribution.
We then fit to each of them separately, derive the form factor from each fit,
and analyze the distribution of these values to generate the error bands
for the form factors and the errors of the extracted radii.
The theoretical errors are estimated by varying the number of
effective vector meson poles. The first error thus gives
the uncertainty due to the fitting procedure (bootstrap) and the data while
the second one reflects the accuracy of the
spectral functions underlying the dispersion-theoretical analysis.
Note that these two errors are not in a strict one-to-one
correspondence to the commonly given statistical and systematic errors.

%

\end{document}